\documentclass[letterpaper,american,twocolumn,superscriptaddress,aps,prb,citeautoscript]{revtex4-1}
\usepackage[T1]{fontenc}
\usepackage[utf8]{inputenc}
\usepackage{babel}
\usepackage{amsmath}
\usepackage{amssymb}
\usepackage{graphicx}
\usepackage[unicode=true,pdfusetitle,
 bookmarks=true,bookmarksnumbered=false,bookmarksopen=false,
 breaklinks=false,pdfborder={0 0 1},backref=false,colorlinks=false]
 {hyperref}

\makeatletter

\pdfpageheight\paperheight
\pdfpagewidth\paperwidth

\providecommand{\tabularnewline}{\\}

\usepackage{diagbox}
\usepackage{makecell}
\usepackage{upgreek}
\usepackage{bbm} % for identity matrix \mathbbm{1}

\makeatother

\begin{document}

\title{}

\title{Four-photon orbital angular momentum entanglement}

\author{B. C. Hiesmayr}

\affiliation{University of Vienna, Faculty of Physics, Boltzmanngasse 5, A-1090
Vienna, Austria}

\author{M. J. A. de Dood}

\affiliation{Huygens--Kamerlingh Onnes Laboratory, Leiden University, P.O. Box
9504, 2300 RA Leiden, The Netherlands}

\author{W. Löffler}

\email{loeffler@physics.leidenuniv.nl}

\affiliation{Huygens--Kamerlingh Onnes Laboratory, Leiden University, P.O. Box
9504, 2300 RA Leiden, The Netherlands}

\begin{abstract}
\textsf{\textbf{}}Quantum entanglement shared between more than
two particles is essential to foundational questions in quantum mechanics,
and upcoming quantum information technologies. So far, up to 14 two-dimensional
qubits have been entangled\cite{monz2011,yao2012eight}, and an open
question remains if one can also demonstrate entanglement of higher-dimensional
discrete properties of more than two particles \cite{pan2012,shalm2012}.
A promising route is the use of the photon orbital angular momentum
(OAM), which enables implementation of novel quantum information protocols
\cite{walborn2006,lanyon2009}, and the study of fundamentally new
quantum states \cite{hiesmayr2013,wieczorek2008}. To date, only two
of such multidimensional particles have been entangled\cite{mair2001}
albeit with ever increasing dimensionality \cite{dada2011,krenn2013d100,salakhutdinov2012}.
Here we use pulsed spontaneous parametric downconversion (SPDC) \cite{lamaslinares2001}
to produce photon quadruplets that are entangled in their OAM, or
transverse-mode degrees of freedom; and witness genuine multipartite
Dicke-type entanglement \cite{guhne2010,huber2010}. Apart from
addressing foundational questions \cite{popescu2014}, this could
find applications in quantum metrology, imaging, and secret sharing
\cite{hillery1999,yu2008ss}. 
\end{abstract}
\maketitle
\textsf{\textbf{}}

Twin photons that are created by SPDC are correlated in several degrees
of freedom and exhibit quantum entanglement. Apart from the well-known
polarization degrees, the photons can also be correlated in their
spatial degrees; this manifests itself in continuous wavevector or
(the Fourier-related) position entanglement \cite{howell2004}. We
can also explore the spatial correlations using transverse paraxial
optical modes, which can be chosen to form a discrete, orthogonal,
and complete set, which is very useful for quantum information applications.
In contrast to the 2-dimensional polarization space, the transverse
mode space is in principle infinite dimensional, only limited by diffraction
and the transverse size of optical components. Recently the use of
computer-controlled holograms enabled steep increase of the single-particle
Hilbert space up to 100 dimensions \cite{dada2011,krenn2013d100}.
An experimentally useful choice of transverse modes are the Laguerre-Gauss
(LG) modes where the azimuthal part factorizes and describes phase
vortices $\exp\left(i\ell\phi\right)$, where $\phi$ is the azimuth
and $\ell=-\infty\dots\infty$ determines the twisting number of the
wavefront; corresponding to an orbital angular momentum of $\ell\hbar$
per photon \cite{allen1992} (in addition to the spin angular momentum).
The LG and the related Hermite-Gauss modes have well-known propagation
dynamics, thus they are suitable for long-distance distribution of
high-dimensional entanglement.

A photon pair produced by SPDC shows quantum entanglement in the transverse-mode
and in particular the OAM degree of freedom \cite{mair2001}, this
has already been used in a large number of quantum mechanical tests.
The quantum correlations can be understood by considering the rotational
symmetry of the SPDC process (on the wavelength scale, the crystal
is rotationally symmetric); due to Noether's theorem, this leads to
conservation of OAM. Therefore, we can write, in analogy to the well-known
wavevector-space SPDC Hamiltonian, the OAM SPDC Hamiltonian in case
of a Gaussian pump beam as

\begin{equation}
H=\sum_{\ell=-\infty}^{\infty}\frac{1}{2}i\kappa\hbar\left(a_{\ell}^{\dagger}a_{\bar{\ell}}^{\dagger}-a_{\ell}a_{\bar{\ell}}\right)\label{eq:hamilton}
\end{equation}

where $a_{\ell}^{\dagger}$ is the creation operator for a photon
with OAM $\ell$, $\bar{\ell}\equiv-\ell$ and $\kappa$ describes
the strength of the nonlinear interaction. A single SPDC photon pair
is produced by the lowest-order term of the series expansion of $|\Psi\rangle=\exp\left(-it/\hbar\,H\right)|0\rangle$,
leading to $|\Psi_{2}\rangle=\gamma\sum_{\ell=1}^{\infty}|1_{\ell};1_{\bar{\ell}}\rangle$
which describes two photons that are perfectly anticorrelated in their
OAM \cite{mair2001,dada2011} ($|n_{k}\rangle$ is a state with $n$
photons in mode $k$). The single-pass amplitude gain $\gamma\propto\kappa t$
depends linearly on the pump beam intensity $I_{p}$. We explore here
the next-order terms corresponding to the simultaneous production
of 2 OAM photon pairs (for $\ell\neq0$ modes):

\begin{equation}
|\Psi_{4}\rangle\propto\gamma^{2}\left(\sum_{i,j=1,i\neq j}^{\infty}|1_{\ell_{i}};1_{\ell_{j}};1_{\bar{\ell}_{i}};1_{\bar{\ell}_{j}}\rangle+2\sum_{\ell=1}^{\infty}|2_{\ell};2_{\bar{\ell}}\rangle\right).\label{eq:psi4}
\end{equation}

\begin{figure}
\includegraphics[width=1\columnwidth]{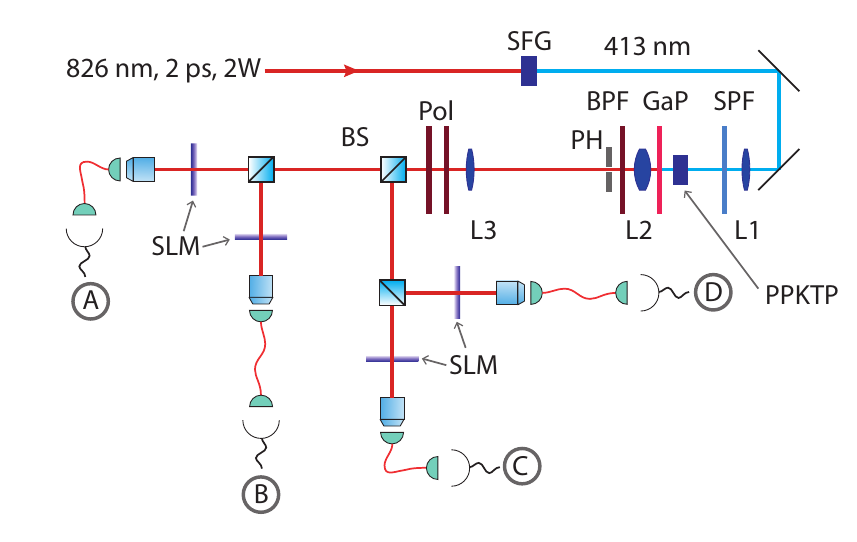}

\caption{\label{fig:expsetup}\textbf{Experimental implementation}. A frequency-doubled
(SFG) mode-locked picosecond laser is short-pass filtered (SPF) and
focussed (L1) into the periodically poled potassium titanyl phosphate
(PPKTP) crystal. The SPDC photons are spectrally filtered with a GaP
plate and a band-pass filter (BPF), and distributed with beam splitters
(BS) to the 4 equal detection units. The crystal facet is imaged with
a telescope (L2 \& L3) onto the spatial light modulators (SLM), which
in turn is far-field imaged onto the core of the detection single
mode fibers connected to single photon counters. The pinhole (PH)
selects the 1st diffraction order of the SLM holograms. We explore
the 4-photon transverse-mode space by changing the holograms on the
SLMs and recording 4-fold coincidence events with a multi-channel
time tagging computer card.}
\end{figure}

This state can be seen as a result of interference in a double-pair
emission process, which for production of multiple photon pairs is
much more stable than interferometric SPDC \cite{lamaslinares2001}
experiments. Because the 4-photon term (Eq.~\ref{eq:psi4}) depends
quadratically on the pump beam intensity, we use a picosecond pulsed
laser, further we use 1~nm wide band pass filters for the downconverted
photons to limit spectral (and temporal) labelling \cite{yorulmaz2014}.
Since we use type-I SPDC, all produced photons have the same polarization.
We use nonpolarizing beamsplitters to separate them, and a combination
of phase-only spatial light modulation and projection onto the core
of single mode fibers to perform projective measurements in transverse-mode
space, see Fig.~\ref{fig:expsetup}. 

\begin{figure}
\includegraphics[width=1\columnwidth]{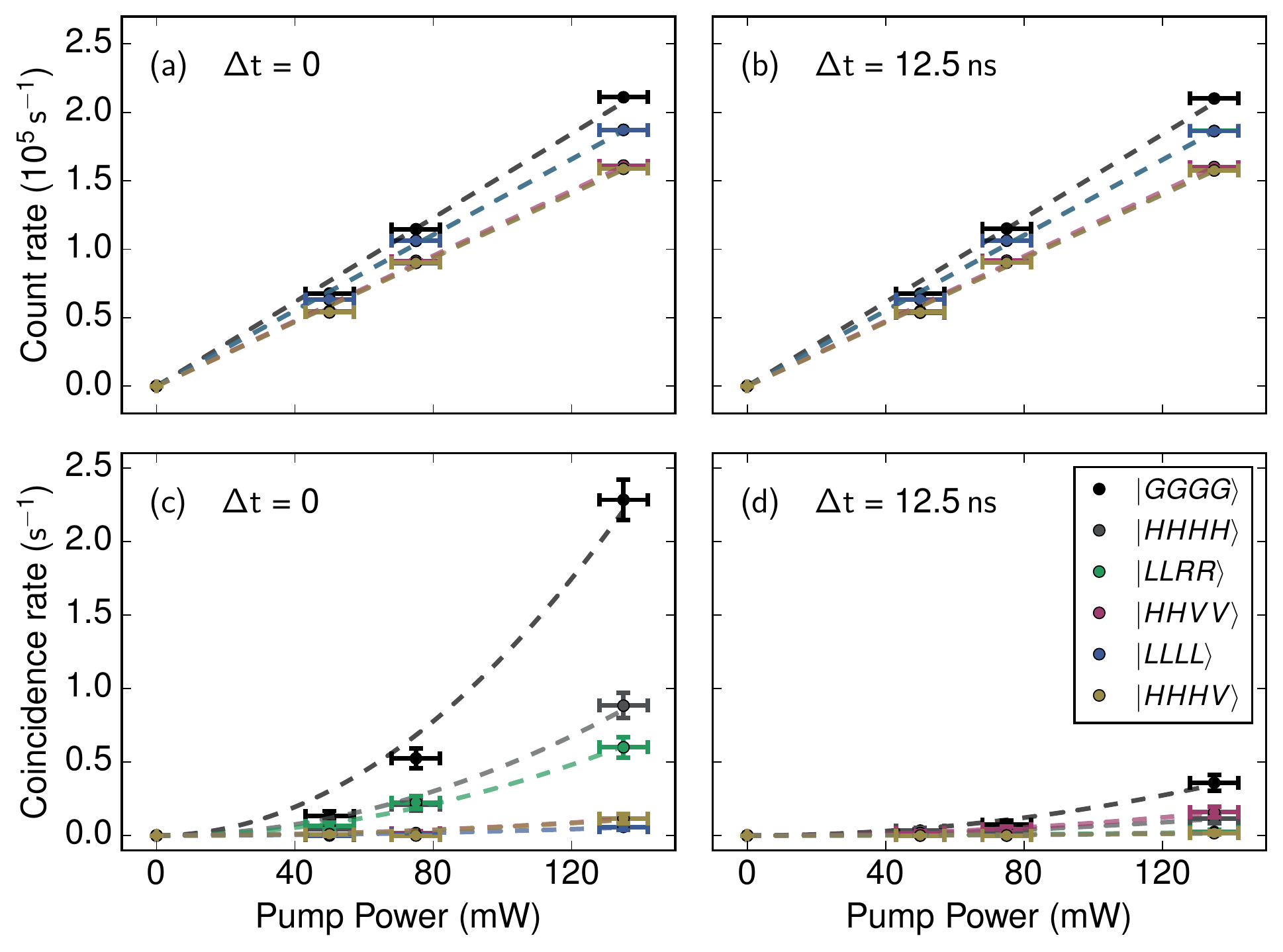}

\caption{\label{fig:powerdep}\textbf{4-photon OAM state production.} Pump-power
dependent single-photon count rate (a, b) and 4-fold coincidence rate
(c, d). The single rates depend linearly on pump power, while the
4-fold coincidences rates show quadratic dependency. The comparison
between zero time delay (a, c) and the case where photon 3 and 4 are
delayed by 12.5~ns shows that most detected 4-photon detection events
are indeed due to 4 photons that were produced within a single pump
pulse. }
\end{figure}

To study the SPDC produced state from Eq.~\ref{eq:psi4}, we record
4-photon correlations for different detection modes, pump powers,
and relative detection times, shown in Fig.~\ref{fig:powerdep}.
We restrict ourselves to the fundamental Gaussian ($G$) mode and
the 2-dimensional first-order mode space. The 3 mutually unbiased
bases of the latter are $\left\{ LG_{+1}^{0},\,LG_{-1}^{1}\right\} \equiv\left\{ R,\,L\right\} $
in the Laguerre-Gauss basis, $\left\{ HG_{0,1},\,HG_{1,0}\right\} \equiv\left\{ H,\,V\right\} $
in the Hermite-Gauss basis, and $\left\{ HG_{0,1}^{45},\,HG_{1,0}^{45}\right\} \equiv\left\{ D,\,A\right\} $
in the $45^{\circ}$ rotated Hermite-Gauss basis, in analogy to the
polarization case \cite{padgett1999}. We designed the experiment
such that the single count rates depend only weakly on detection mode
(Fig.~\ref{fig:powerdep} a, b). The 4-fold coincidence rate $\Gamma$
for the case that all photons are produced by the same pump pulse
(Fig.~\ref{fig:powerdep} c) is highest if all 4 photons are projected
onto the fundamental Gaussian mode ($|GGGG\rangle$), and depends
quadratically on the pump power as expected. From the ratio between
(d) and (c), whereas in (d) detectors (A, B) and (C, D) are set to
detect photons produced in two different laser pulses, we can estimate
the fraction of uncorrelated events to be 10\% ($\Gamma^{\Delta t=12.5\mathrm{ns}}/\Gamma^{\Delta t=0})$.
This is similar to the fraction of ``forbidden'' events (e.g., $\Gamma_{HHHV}/\Gamma_{HHHH}\approx0.1$).
We think the latter occur due to experimental imperfections; in contrast
to polarization experiments we require here mode-matching between
all 4 detectors and the pump beam simultaneously. This argument is
supported by the fact that both ratios are largely independent on
pump power, suggesting that contributions from higher-order multiphoton
states are low.

These results support an intuitive explanation of the structure of
Eq.~\ref{eq:psi4}, keeping in mind that the photons are produced
in pairs: The first term in Eq.~\ref{eq:psi4} that contains photons
with different OAM $|\ell|$ occurs if the second pair is spontaneously
emitted and uncorrelated to the first pair, while the second term
corresponds to the case where the second pair is produced by a stimulated
process giving a perfect clone of the first pair \cite{ou2005,torren2012,riedmatten2004}.
The different $|\ell|$ values in the first term in Eq.~\ref{eq:psi4}
is then simply a consequence of forbidden perfect quantum cloning,
both terms together demonstrate the possibility of optimal quantum
cloning in stimulated SPDC\cite{simon2000}. 

\begin{figure}
\begin{tabular*}{0.98\columnwidth}{@{\extracolsep{\fill}}lcc}
\includegraphics[width=0.3\columnwidth]{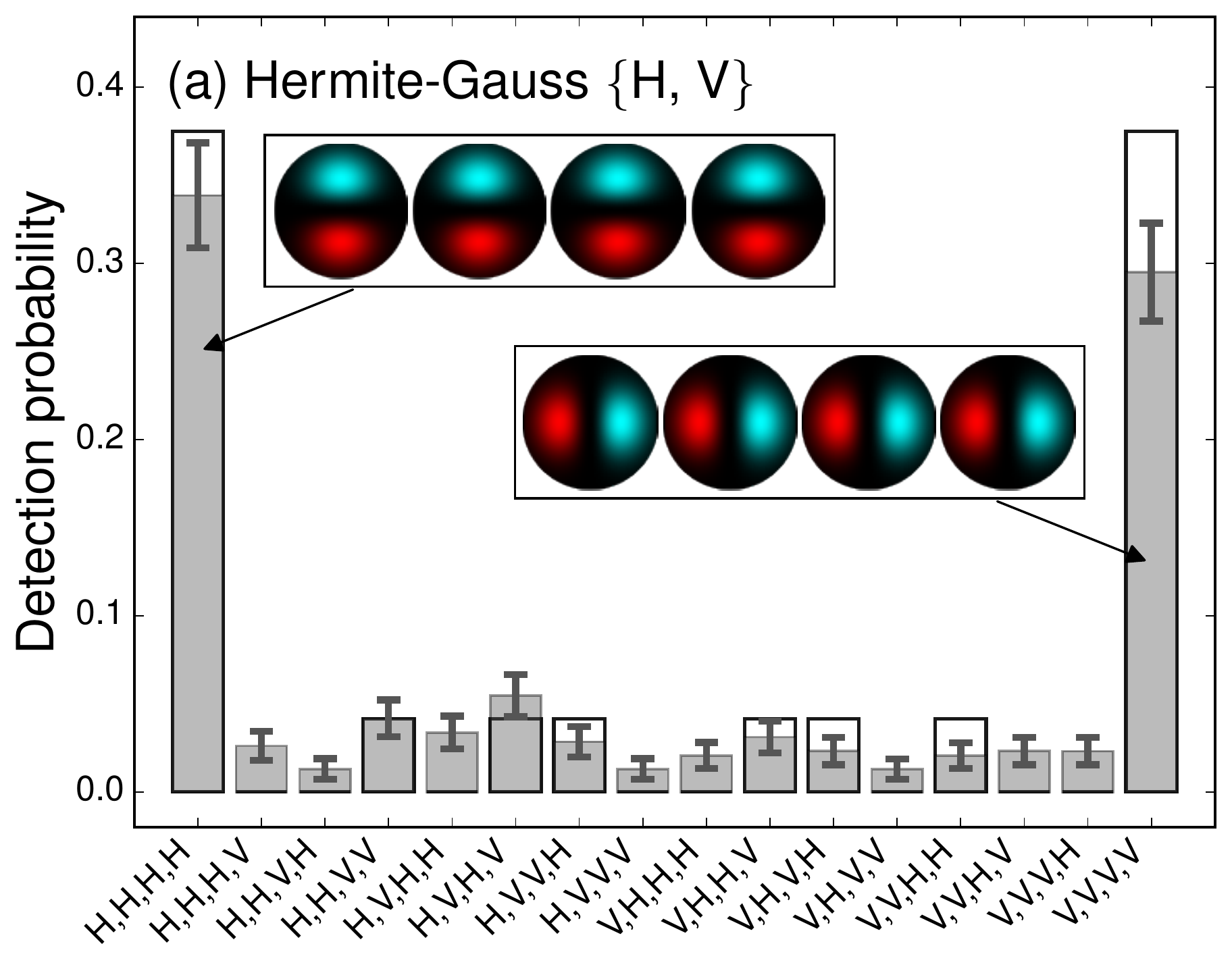} & \includegraphics[width=0.3\columnwidth]{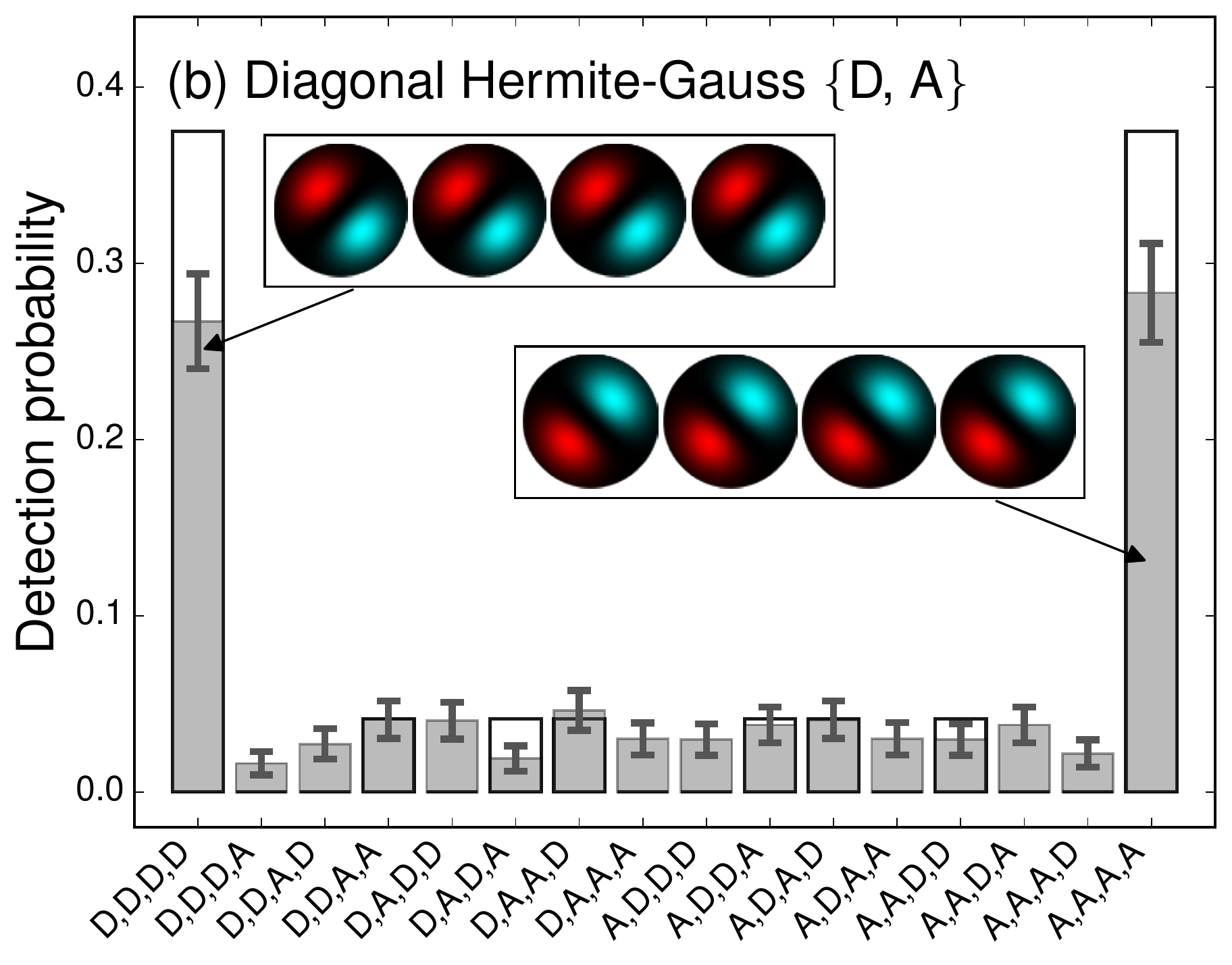} & \includegraphics[width=0.3\columnwidth]{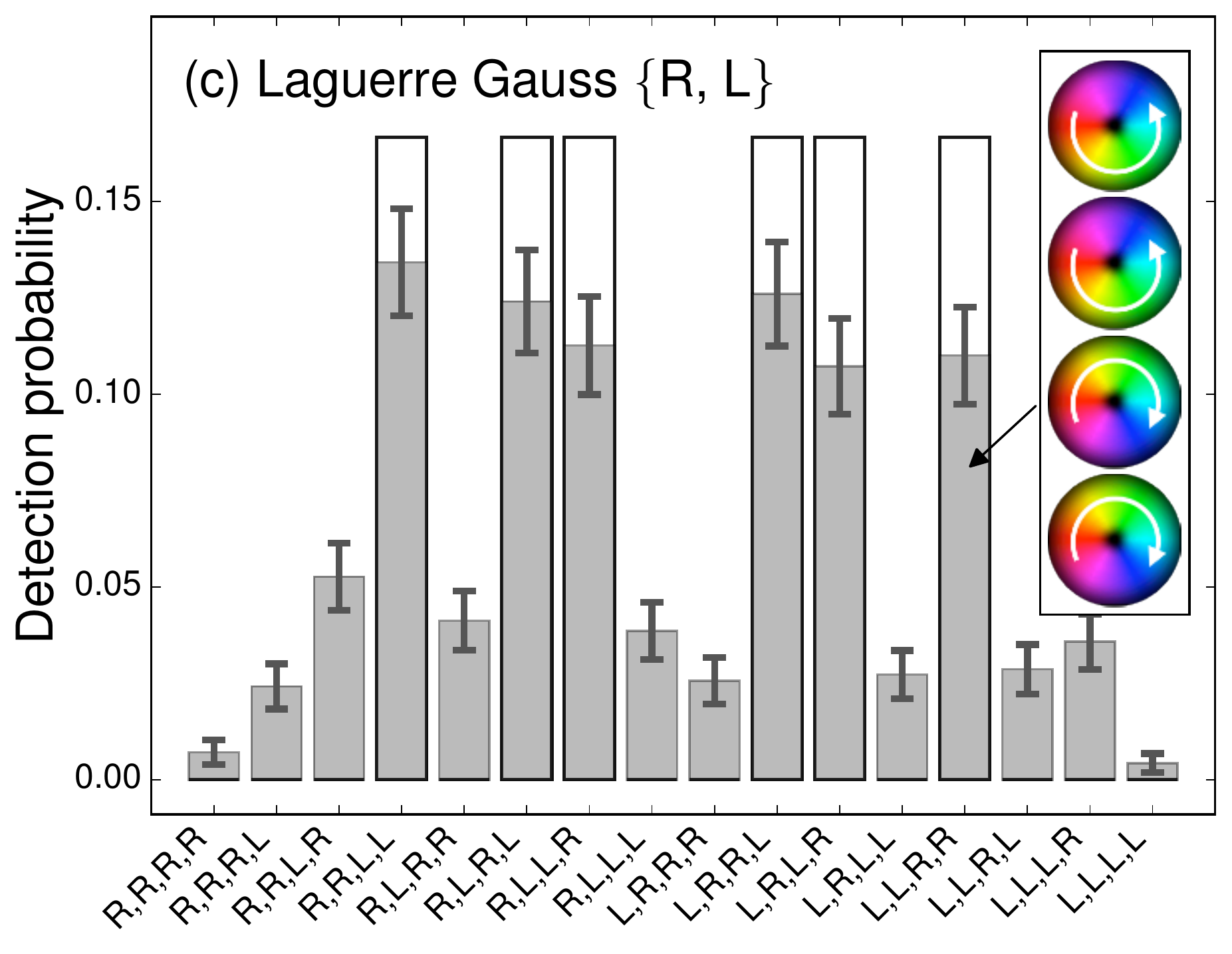}\tabularnewline
\end{tabular*}

\caption{\label{fig:4phcorrs}\textbf{4-photon OAM correlations.} 4-photon
joint detection probabilities in each of the 3 mutually unbiased bases:
Hermite-Gauss $\{H,\,V\}$, diagonal Hermite-Gauss $\{D,\,A\}$, and
Laguerre Gauss $\{R,\,L\}$. Labels indicate the phase and amplitude
structure of the detected spatial modes. Gray bars: experimental data,
error bars show statistical errors. Boxes: theory. Experimental integration
time per data point was 24 minutes. The probabilities are normalized
to unity within each basis.}
\end{figure}

Now we study the quantum correlations of the SPDC-produced 4-photon
state, for which we focus on the 2-dimensional first-order mode space
with OAM $\ell=\pm1$, and record 4-fold coincidences for all detector
mode combinations in each basis. Fig.~\ref{fig:4phcorrs} compares
experimental results and theoretical prediction for the SPDC-produced
state, which becomes in this case (Eq.~\ref{eq:psi4}):

\begin{equation}
|\Psi_{4}^{(2)}\rangle\propto|2_{\ell};2_{\bar{\ell}}\rangle\label{eq:dicke}
\end{equation}

with $\ell=1$. Note that Eq.~\ref{eq:dicke} is valid for any $\ell$,
if we limit our detection to $\pm\ell$ modes, a 2D Hilbert space
per photon. Experimentally, the beamsplitters generate all possible
permutations of photons (Fig.~\ref{fig:expsetup}), so $|\Psi_{4}^{(2)}\rangle$
becomes in the detector-basis 
\begin{equation}
|D_{4}^{(2)}\rangle\propto|\ell\ell\bar{\ell}\bar{\ell}\rangle+|\ell\bar{\ell}\ell\bar{\ell}\rangle+|\bar{\ell}\ell\ell\bar{\ell}\rangle+|\ell\bar{\ell}\bar{\ell}\ell\rangle+|\bar{\ell}\ell\bar{\ell}\ell\rangle+|\bar{\ell}\bar{\ell}\ell\ell\rangle\label{eq:dickedetected}
\end{equation}
which is the symmetric Dicke state of $N=4$ photons with $N/2=2$
excitations. This state is in particular interesting as it is robust
to photon losses and has the largest distance from not genuine multipartite
entangled states \cite{toth2007}. Entanglement in such states can
easily be verified by measuring the total collective spin along the
$x,y$ directions \cite{toth2007}; if $\langle\mathcal{W}_{4}^{(2)}\rangle=\langle J_{x}^{2}\rangle+\langle J_{y}^{2}\rangle>5$
(for $N=4$; $J_{i}=\frac{1}{2}\sum_{k}\sigma_{i}^{(k)}$ where, e.g.,
$\sigma_{y}^{(2)}=\mathbbm{1}\otimes\sigma_{y}\otimes\mathbbm{1}\otimes\mathbbm{1}$),
the state is non-separable, i.e., entangled. We obtain experimentally
$\langle\mathcal{W}_{4}^{(2)}\rangle=5.17\pm0.09$, thus verifying
entanglement in the 4-photon OAM state.

$\langle\mathcal{W}_{4}^{(2)}\rangle$ can also be used to detect
genuine multipartite entanglement \cite{guhne2010,huber2010,bourennane2004},
if it violates $\langle\mathcal{W}_{4}^{(2)}\rangle\leq7/2+\sqrt{3}\approx5.23$
\cite{toth2007}. The proposed theoretical state in Eq.~\ref{eq:dicke},
for which $\langle\mathcal{W}_{4}^{(2)}\rangle=6$, is indeed genuine
multipartite entangled, but our experimental result does not violate
this bound. We argue that experimental imperfections are responsible:
Apart from spectral-filtering issues, we have here the extreme requirement
that all 4 detectors have to be mode matched simultaneously. In contrast
to experiments on polarization entanglement, here, even small misalignment
does not only reduce count rates but also alters the measurement projectors
by inducing small rotations in the respective single-particle Hilbert
space, and the 4-fold mode-matching exponentially amplifies any misalignment.
The experimentally obtained numerical values should therefore be seen
as a lower limit only. We can correct for this partially if we have
access to the full density matrix, as we show now.

\begin{figure}
\includegraphics[width=1\columnwidth]{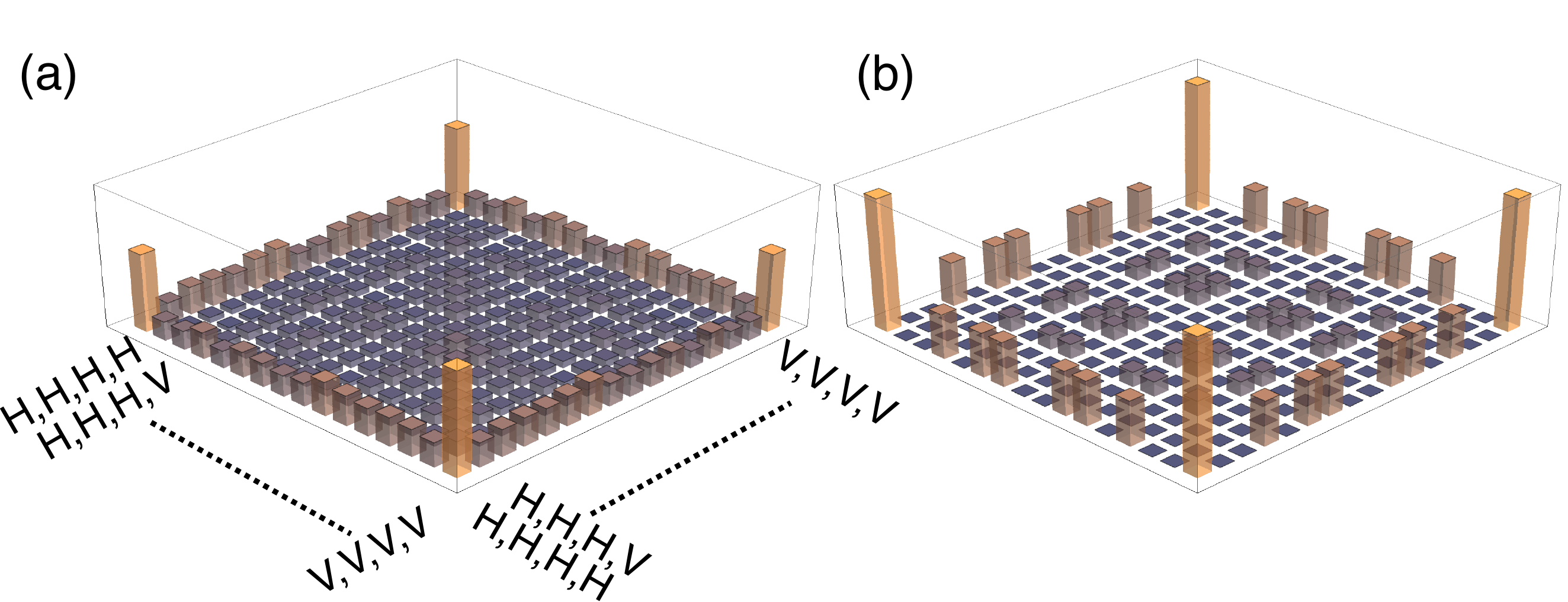}

\caption{\label{fig:rho}\textbf{The reconstructed OAM quantum state.} Modulus
of the reconstructed experimental (a) and theoretical (b) density
matrix (vertical axis scale: $\left[0\dots0.3\right]$). The discrepancy
between experiment and theory is mainly caused by residual misalignment
of the transverse-mode detectors.}
\end{figure}

We use tomographic quantum state reconstruction \cite{james2001}
to obtain the most likely density matrix $\rho$ describing the detected
state. Experimental integration time is 120~s for each detector setting,
and we additionally combine the data with the more accurate measurements
of the basis-state correlations shown in Fig.~\ref{fig:4phcorrs}.
In Fig.~\ref{fig:rho} we compare the resulting experimental density
matrix to the theoretical expectation (Eq.~\ref{eq:dickedetected}).
As a multipartite entanglement witness we use the one proposed in
Ref.~\onlinecite{huber2011}, which is also more resilient against
noise compared to the collective-spin witness above. Violation of
the inequality $I_{2}^{4}[\rho]\leq0$ indicates genuine multipartite
entanglement of $\rho$. This witness is constructed such that it
is optimal for the Dicke state with two excitations $|D_{4}^{(2)}\rangle$
and gives 1 in this case. For analysis of the experimental density
matrix, we apply local single-qubit rotations and search for maximum
violation of the witness (see methods). For 3 independently measured
data sets, we obtain $I_{2}^{4}[\rho]=\left\{ 0.28,\,0.32,\,0.34\right\} $,
thus giving strong indication of genuine 4-photon OAM entanglement.

We have studied here the zero- and first-order modes with $\ell=\{0,\,\pm1\}$.
Exploration of the rich correlations in the full higher-dimensional
multipartite state (Eq.~\ref{eq:psi4}) will require higher pump
beam intensities and therefore the use of nonlinear crystals with
higher damage threshold, both within reach today. The 4-photon OAM
entangled state that we have produced and characterized here might
open new possibilities and protocols in multi-party quantum secret
sharing \cite{hillery1999} with Dicke states in the sense that here,
more information per photon quadruplet can be exchanged or the security
increased by the high-dimensional nature of the OAM or transverse-mode
degrees of freedom \cite{yu2008ss}. Further, the spatial correlations
carried by our multi-photon states might enable new options in quantum
metrology, microscopy, and imaging.

\textbf{Appendix}

\emph{Photon quadruplet creation in high-gain SPDC:}

We use a PPKTP crystal cut for degenerate collinear type-I SPDC. The
photon pairs at 826~nm experience a different group index than the
pump beam of 413~nm in PPKTP, therefore temporal labelling can occur.
The group index difference is $n_{g}(2\omega)-n_{g}(\omega)=0.456$
at 300~K, corresponding to a group velocity dispersion of $D=1.5$~ps/mm,
and a group velocity walk-off length $L_{gv}=1.3$~mm for $\Delta t=2$~ps
pulses. Thus, we choose a 1~mm long crystal. We separate the pump
beam with an anti reflection coated GaP plate and select a narrow
frequency window around the degenerate wavelength of 826~nm with
a 1~nm wide spectral filter.

The number of available entangled transverse modes is determined by
the pump beam size in the crystal, at perfect phase matching. To achieve
sufficient count rates, we focus the pump beam to $50\,\mu$m (while
the detection mode waist in the crystal is 100~$\mu$m), which results
in $\sim9$ transverse modes.

If not noted otherwise, we use a pump power of 70~mW, which corresponds
to a peak intensity at the focus of 11.2~kW cm$^{-2}$. This is close
to the PPKTP damage threshold; above this, gray-tracking was observed.
Elevated temperatures or other materials, such as periodically poled
Lithium Niobate, would enable the use of higher intensities.

\emph{Entanglement witness optimization:}

Local unitary transformations, complex basis rotations, are allowed
to be applied to each part of a multipartite state without changing
its entanglement properties. Most entanglement witnesses, including
the witness for $n$ particles $I_{m}^{n}[\rho]$ optimized for a
Dicke state with $m$ excitations from Ref.~\cite{huber2011}, only
detect entanglement optimally in a particular basis, and optimization
over basis rotations has to be done. Here, we optimize $I_{m}^{n}[\rho]$
by transformation of the experimentally determined (or theoretical)
density matrix $\rho$ as follows: $\rho\rightarrow U\rho\,U^{\dagger}$,
where $U=U_{1}\otimes U_{2}\otimes U_{3}\otimes U_{4}$ with the generic
unitary

\[
U_{i}=\begin{bmatrix}\exp i\alpha_{i}\cos\gamma_{i} & \exp i\beta_{i}\sin\gamma_{i}\\
-\exp i(\alpha_{i}-\delta_{i})\sin\gamma_{i} & \exp i(\beta_{i}-\delta_{i})\cos\gamma_{i}
\end{bmatrix}
\]

for the real parameters $\left\{ \alpha_{i},\,\beta_{i},\,\delta_{i},\,\gamma_{i}\right\} $.
These 16 parameters are simultaneously optimized using robust unconstrained
numerical optimization routines.

\textbf{Acknowledgements} We thank Martin van Exter for invaluable
discussions. M.D. and W.L. acknowledge support from NWO and FOM. B.H.
acknowledges support from the Austrian Science Fund (FWF 23627).

\bibliographystyle{naturemagw}
\bibliography{bibliography}

\end{document}